\documentclass[twocolumn]{aastex631}

\usepackage{amsmath}

\shorttitle{MAPS-ing the Planet Wake in HD163296}
\shortauthors{Calcino et al.}
\graphicspath{{./}{figures/}}

\begin{document}



\title[MAPS-ing the Planet Wake]{Mapping the Planetary Wake in HD~163296 with Kinematics}



\author[0000-0001-7764-3627]{Josh Calcino}
\altaffiliation{jcalcino@lanl.gov}
\affiliation{Theoretical Division, Los Alamos National Laboratory, Los Alamos, NM 87545, USA}

\author[0000-0001-7641-5235]{Thomas Hilder}
\affiliation{School of Physics and Astronomy, Monash University, Clayton, Vic 3800, Australia}

\author[0000-0002-4716-4235]{Daniel J. Price}
\affiliation{School of Physics and Astronomy, Monash University, Clayton, Vic 3800, Australia}

\author[0000-0001-5907-5179]{Christophe Pinte}
\affiliation{School of Physics and Astronomy, Monash University, Clayton, Vic 3800, Australia}
\affiliation{Université Grenoble Alpes, CNRS, IPAG, F-38000 Grenoble, France}

\author{Francesco Bollati}
\affiliation{Dipartimento di Scienza e Alta Tecnologia, Universit\`a degli Studi dell'Insubria, Via Valleggio 11, I-22100, Como, Italy}

\author[0000-0002-2357-7692]{Giuseppe Lodato}
\affiliation{Dipartimento di Fisica, Universit\`a degli Studi di Milano, Via Celoria 16 I-20133 Milano, Italy}

\author[0000-0001-5898-2420]{Brodie J. Norfolk}
\affiliation{Centre for Astrophysics and Supercomputing (CAS), Swinburne University of Technology, Hawthorn, Victoria 3122, Australia}




\defcitealias{bollati2021}{B21}

\begin{abstract}

We map the planetary wake associated with the embedded protoplanet creating the CO kink in the disk of HD~163296. We show that the wake can be traced by a series of correlated perturbations in the peak velocity map. The sign change of the perturbations across the disk major axis confirm that the wake induces predominantly radial motion, as predicted by models of planet-disk interaction. These results provide the first direct confirmation of planet wakes generated by Lindblad resonances. Mapping the wake provides a constraint on the disk aspect ratio, which is required to measure the mass of the planet.
\end{abstract}



\section{Introduction} \label{sec:intro}

The disk around HD~163296 displays the kinematic signatures associated with an embedded planet \cite[][]{pinte2018}, which was first hypothesized by \citet{grady2000}. Higher spatial resolution observations have revealed additional substructure in mm-continuum and CO line observations from the Disk Substructures At High Angular Resolution Project (DSHARP) \citep{andrews2018, jhuang2018, pinte2020, discminer} and more recently the Molecules with ALMA at Planet forming Scales (MAPS) \citep{oberg2021, teague2021} large programs using the Atacama Large Millimetre/submillimeter Array (ALMA). \cite{teague2021} pointed out spiral features both in channel maps and velocity residuals. Velocity kinks arise in the CO channel maps due to perturbations in the motion of gas, which modifies the projected velocity. We define a `velocity kink' as a localised distortion of the isovelocity curve, caused by deviations from Keplerian motion which shift the line emission into an adjacent channel. They can be induced by, for example, spiral arms, gravitating bodies, or gravitational instability in the disk. In particular, velocity kinks detected in the CO emission were associated with an embedded planet \citep{pinte2018}. 

In \cite{bollati2021} (hereafter \citetalias{bollati2021}), we recently developed a semi-analytic theory of velocity kinks caused by planet-disk interaction, based on the mathematical theory of planet-induced density waves launched at Lindblad resonances developed by \citet{goldreich1979,goldreich1980,goodman2001,ogilvie2002} and \citet{rafikov2002}. Applying this model to the kinematic detection of a planet in HD~163296 made by \cite{pinte2018} allows one to interpret velocity kinks as occurring whenever the spiral wake from the planet intersects a velocity channel in the data. More puzzling was our finding in \citetalias{bollati2021} that such kinks should therefore extend throughout the disk, including far from the planet (see Figure 6 of \citetalias{bollati2021}). Hence, non-localised `secondary' kinks away from the planet should be detectable. In \citetalias{bollati2021} we tried to explain away the apparent conflict between the non-detection of these secondary kinks in the data --- perhaps being washed out by the finite beam, or suppressed due to viscous or other form of damping occurring in the disk. With the MAPS data, the former hypothesis is no longer tenable.

In this Letter, we demonstrate that not only are secondary kinks observed in HD~163296, they can be used to map the wake generated by the embedded planet found by \citet{pinte2018} and \citet{teague2018} over a large fraction of the disk. In particular, we show that the spiral wake generated by the planet can be easily traced in a peak velocity map given sufficient spatial and spectral resolution. By determining the shape of the planet wake we constrain the aspect ratio of the disk, which is needed in order to measure the planet mass from observed velocity kinks \citepalias{bollati2021}.

\section{Methods}

\subsection{Observations}

We used $^{12}$CO (2-1) \texttt{robust=0.5} line emission observations of HD~163296 from the MAPS large program (2018.1.01055.L, \citealt{MAPSI2021, MAPSII2021})\footnote{The data are available for download at \url{http://alma-maps.info/}.}. The observations have a JvM correction \citep{jvm1995} which accounts for a mismatch between the CLEAN and dirty beam sizes when a CLEAN model image is combined with the residuals.
The observations have a beam size of approximately $0.14\arcsec \times 0.11 \arcsec$ with a position angle of $104^\circ$ and spectral resolution of 92 ms$^{-1}$ and a spacing of 200 ms$^{-1}$. These are the same observations used by \cite{teague2021} in their kinematic analysis of HD~163296. We assumed a line-of-sight velocity of $v_{\textrm{los}}= 5.76$ km/s, as found by \cite{teague2021}. The emitting layer of CO was estimated by \cite{law2021} using the method developed by \cite{pinte2018b}. The resulting surface layer was fitted by a flared disk structure with an exponential taper, with the following functional form
\begin{equation}
    z(r) = z_0 \left( \frac{r}{1\arcsec} \right) ^\phi \exp \left( -\left[ \frac{r}{r_\textrm{taper}} \right]^{\psi} \right).
    \label{eq:height}
\end{equation}
We adopted the best fit values of HD~163296 from \cite{law2021} of $z_0 = 0.388$, $\phi = 1.851$, $r_\textrm{taper} = 2.362\arcsec$, and $\psi = 1.182$ when computing the CO emitting layer.

\subsection{Hydrodynamical Simulations and Radiative Transfer}

We used the smoothed particle hydrodynamics (SPH) code {\sc{phantom}} \citep{phantom2018} to simulate the interaction of a 3 M$_\textrm{J}$ planet with a gas-only disk in 3D. We employed $2\times 10^{6}$ SPH particles to model the gas disk with mass $10^{-2}$ M$_\odot$ orbiting a 1.9 M$_\odot$ central star. Both the planet and central star were modeled using sink particles \citep{bate1995}. The central star was given an accretion radius of 5 au, while the planet has a radius 5.56 au, or 0.25 the Hill radius. The SPH gas particles were initialized with a pressure corrected Keplerian velocity, assuming a power-law distribution with an exponential taper given by
\begin{equation}
    \Sigma (r) \propto (r/r_\textrm{p,i})^{-p} \times \exp [-(r/r_c)^{2-p}],
\end{equation}
where $r_\textrm{p,i} = 280$ au is the initial location of the planet, $p = 1$, and $r_c = 100 $ au. The temperature profile was assumed to be vertically isothermal with the sound speed $c_\textrm{s} \propto (r/r_\textrm{p,i})^{-q}$, where we set $q = 0.35$. The preceding parameters were selected following the model from \citet{pinte2018}, except for the initial planet location which was adjusted to allow for migration. We assumed a disk aspect ratio $(h/r)_{\textrm{p,i}} = 0.1$ following \citet{deGregorioMonsalvo2013} and \citet{pinte2018}. The disk was allowed to evolve for 50 orbits of the planetary companion, at which point it had migrated to a radius of $r_p = 256$ au from the central star where $h_p/r_p \sim 0.09$, and has a mass $\sim 4$ M$_\textrm{J}$. This difference in mass compared to the analytics is not significant, being smaller than the measurement error from kinematics at present \citep[e.g.][]{pinte2018}.

The output of our SPH simulation was used to generate a Voronoi mesh for input in to the Monte Carlo radiative transfer code {\sc{mcfost}} \citep{pinte2006, pinte2009}. Since we did not include dust in our simulation we assumed that the dust followed the gas particles with a power-law grain size distribution $dn/ds \propto s^{-3.5}$ for $0.03\mu$m $\leq s \leq 1$mm 
with a gas-to-dust ratio of 100. We assumed spherical and homogeneous grains composed of astronomical silicate \citep{weingartner2001}. The central star is set with an effective temperature $T_\textrm{eff} = 9250$ K and a radius $R_\star = 1.6 $ R$_\odot$, giving a luminosity of $\sim 17$ L$_\odot$ \citep{Setterholm2018}.

The temperature profile and specific intensities were generated using $10^8$  photon packets. Images were produced by ray-tracing the computed source function. Our CO isotopologue observations were generated assuming $T_\textrm{gas} = T_\textrm{dust}$ and all molecules were assumed to be in Local Themodynamical Equilibrium (LTE). We assumed an initially constant CO abundance, with an abundance ratio of $^{12}$CO/H$_2 = 5\times 10^{-5}$. This abundance was then altered due to CO freeze out ($T = 20$ K, depletion factor of $10^{-4}$), photo-desorption, and photo-dissociation, following Appendix B of \cite{pinte2018b}.

Channels for the CO emission are created at a separation of 20 ms$^{-1}$. We linearly interpolate over 5 channels and average the interpolated data after weighting by a Hann window function, producing a channel width of 100 ms$^{-1}$ with a spacing of 200 ms$^{-1}$. The images were then smoothed with a Gaussian beam with a size $0.15\times 0.15$ arcseconds. 

\subsection{Spiral Wake Mapping}

In a Keplerian power-law disc, $c_{\rm s} (r) = c_{\textrm{s}, \textrm{p}} (r/r_\textrm{p})^{-q}$, $\Sigma = \Sigma_p (r/r_\textrm{p})^{-p}$ and $\Omega_K = (G M_\star / r^3)^{1/2}$, where $r_\textrm{p}$ is the planet location, $c_{\rm s}$ is the sound speed and $\Sigma$ is the column density, the planetary wake generated by a planet lies on the curve given by \citep{rafikov2002}
\begin{align}
    \varphi _\textrm{wake}(r) = \varphi _\textrm{p} + &\textrm{sign}(r-r_\textrm{p}) \biggl( \frac{h_\textrm{p}}{r_\textrm{p}}\biggr)^{-1}\biggl[ \frac{(r/r_\textrm{p})^{q - 1/2}}{q - 1/2} \nonumber \\
    &-\frac{(r/r_\textrm{p})^{q + 1}}{q + 1}-\frac{3}{(2q - 1)(q + 1)} \biggr],
    \label{eq:wake}
\end{align}
where $(h_\textrm{p}/ r_\textrm{p})$ is the scale height at the planet location. We initially assumed $q=0.35$ and $h_{\textrm{p}}/r_\textrm{p}=0.09$ as in the simulations, but found the outer wake matched better to the prominent kink in the outer disk with a slightly lower $h_{\textrm{p}}/r_\textrm{p}$ of 0.08. 

We followed the prescription outlined in \citetalias{bollati2021} to compute the velocity perturbations created by the planet wake in the disk mid-plane using the non-linear evolution theory, with the same disk and planet parameters as above. We generated these perturbations on a $500\times500$ Cartesian grid and added them to an unperturbed background given by \cite{takeuchi2002}
\begin{equation}
    v_\phi \left(r \right) = \left( \frac{GM_*}{r} \right)^{\frac{1}{2}} \left[ 1 - (p+q) \left( \frac{h_p}{r_p} \right)^2 \left( \frac{r}{r_p} \right)^{1-2q} \right]^{\frac{1}{2}}, \label{eqn:vphi_background}
\end{equation}
where the deviation from Keplerian rotation is due to the gas pressure support. 

We projected the wake and associated velocity fields (Eq. \ref{eq:wake}) onto the CO emitting layer (Eq. \ref{eq:height}) assuming no vertical dependence in the velocity perturbations (see Section~\ref{sec:limitations}), and then onto the plane of the sky to derive a line of sight velocity map of the disk upper surface (using a disk inclination of $i=46^{\circ}$ and position angle PA$\,=313^{\circ}$). In this projection we implicitly assume identical radial temperature profiles in both the mid-plane and the emitting layer. In reality the temperature is vertically stratified, which may result in spirals with a larger pitch angle (see Section~\ref{sec:limitations}).

\begin{figure*}
\centering
\includegraphics[width=1\textwidth]{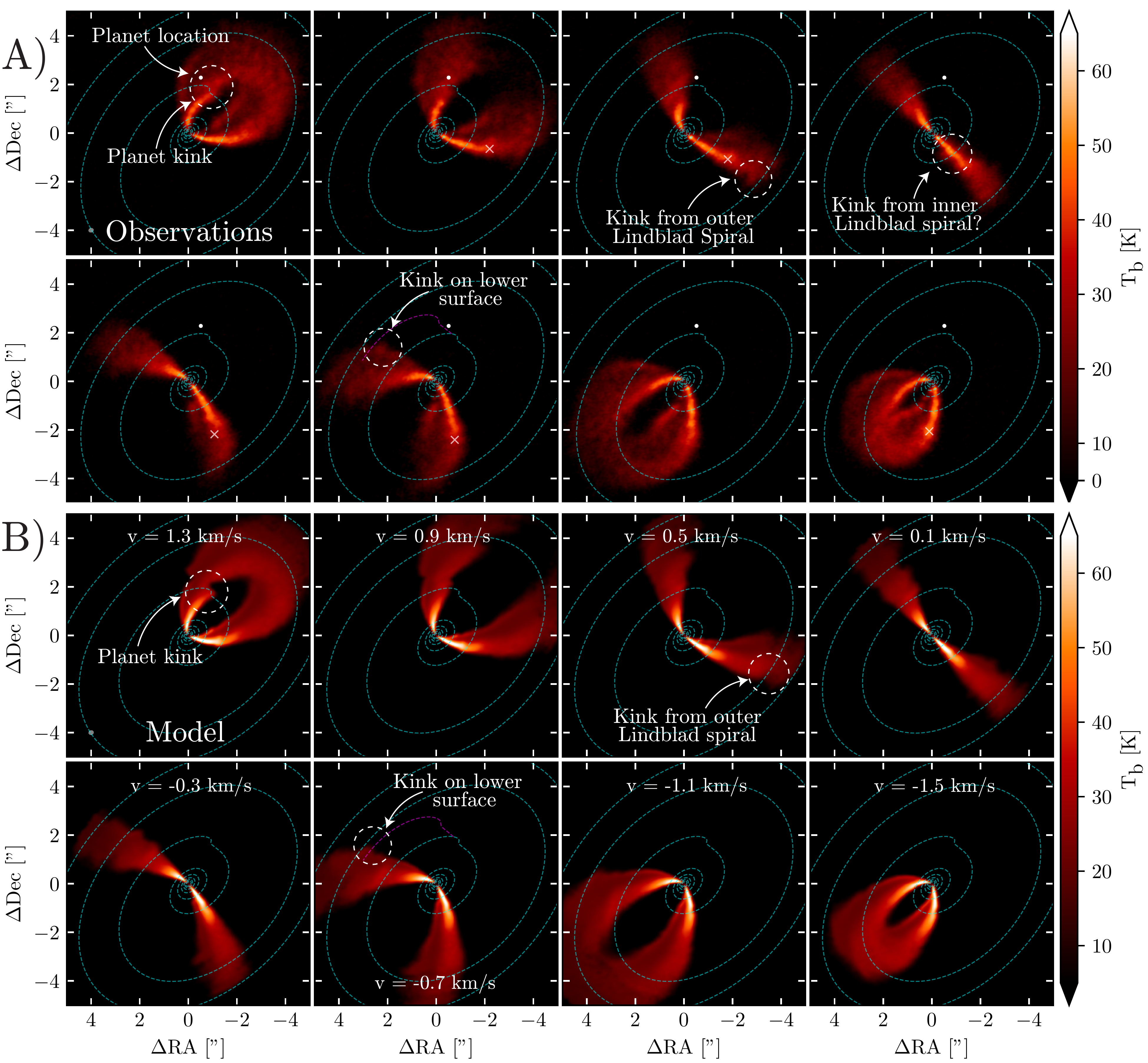}
\caption{Secondary kinks seen in channel maps of $^{12}$CO (2-1) line emission in HD163296 \protect{\citep{oberg2021}} (Subplot~A), along with the same channels from our SPH simulation (Subplot~B). Labels indicate the location of the embedded planet as well as secondary kinks caused by the spiral wake, denoted by the cyan dashed line (Equation~\ref{eq:wake}; projected to the CO emitting layer). The predicted location of the wake explains the main spiral features seen in the channel maps. The magenta line in the second panel on the bottom of each subplot is the wake projected on to the lower surface. The opaque crosses mark the locations of additional kinks that do not appear associated with the planetary wake.}
\label{fig:channels}
\end{figure*}

\begin{figure*}
\centering
\includegraphics[width=0.8\textwidth]{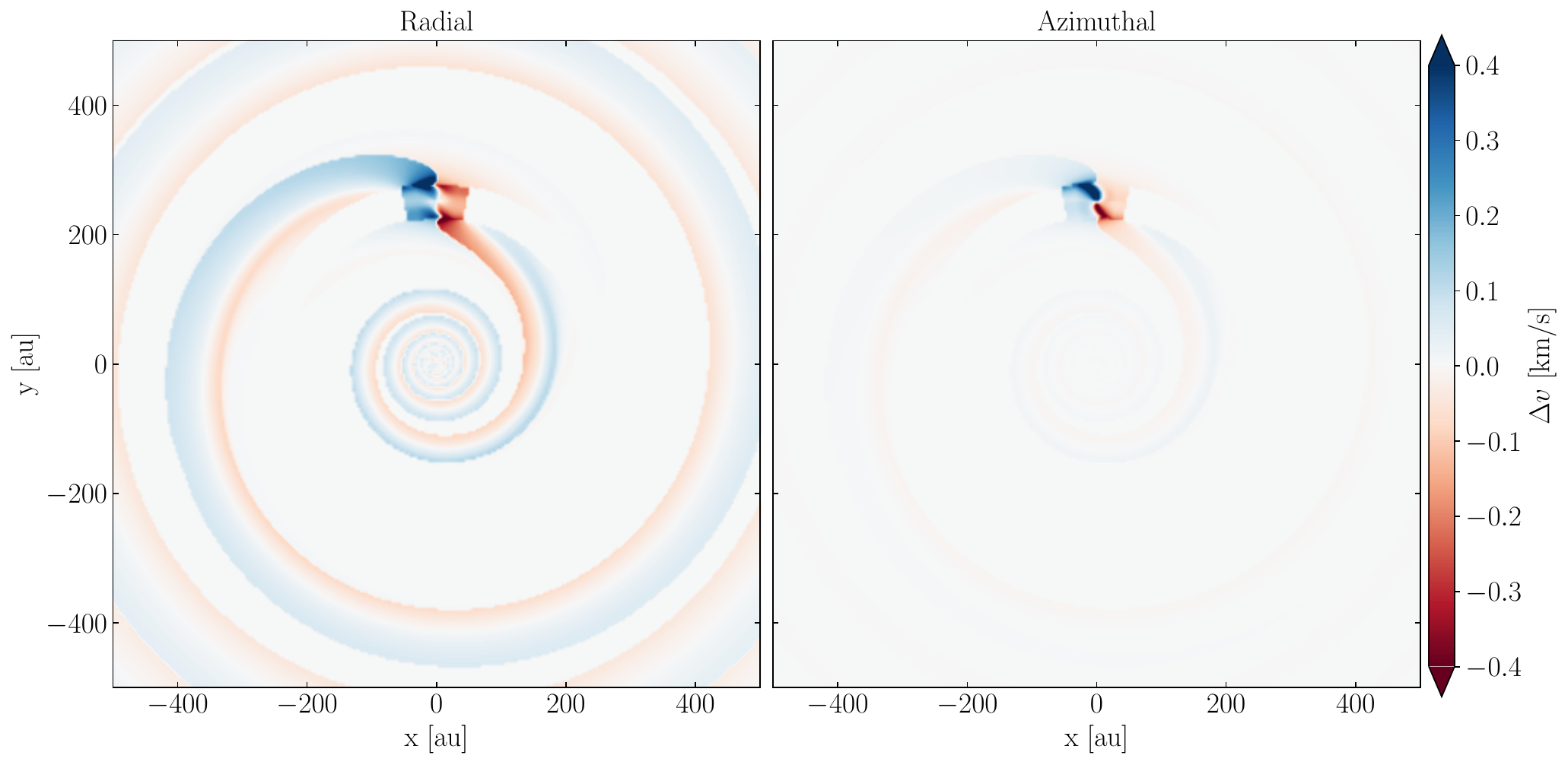}
\caption{Analytic prediction of the velocity perturbation generated by the planetary wake of a 3 M$_\textrm{J}$ planet located at 250 au. Left and Right panels show radial and azimuthal velocity perturbations, respectively. Note that positive radial velocities indicate motion away from the central star, and positive azimuthal velocities indicate super-Keplerian flow.}
\label{fig:analytics}
\end{figure*}

\begin{figure*}
\centering
\hspace{-1cm}
\includegraphics[width=0.8\textwidth]{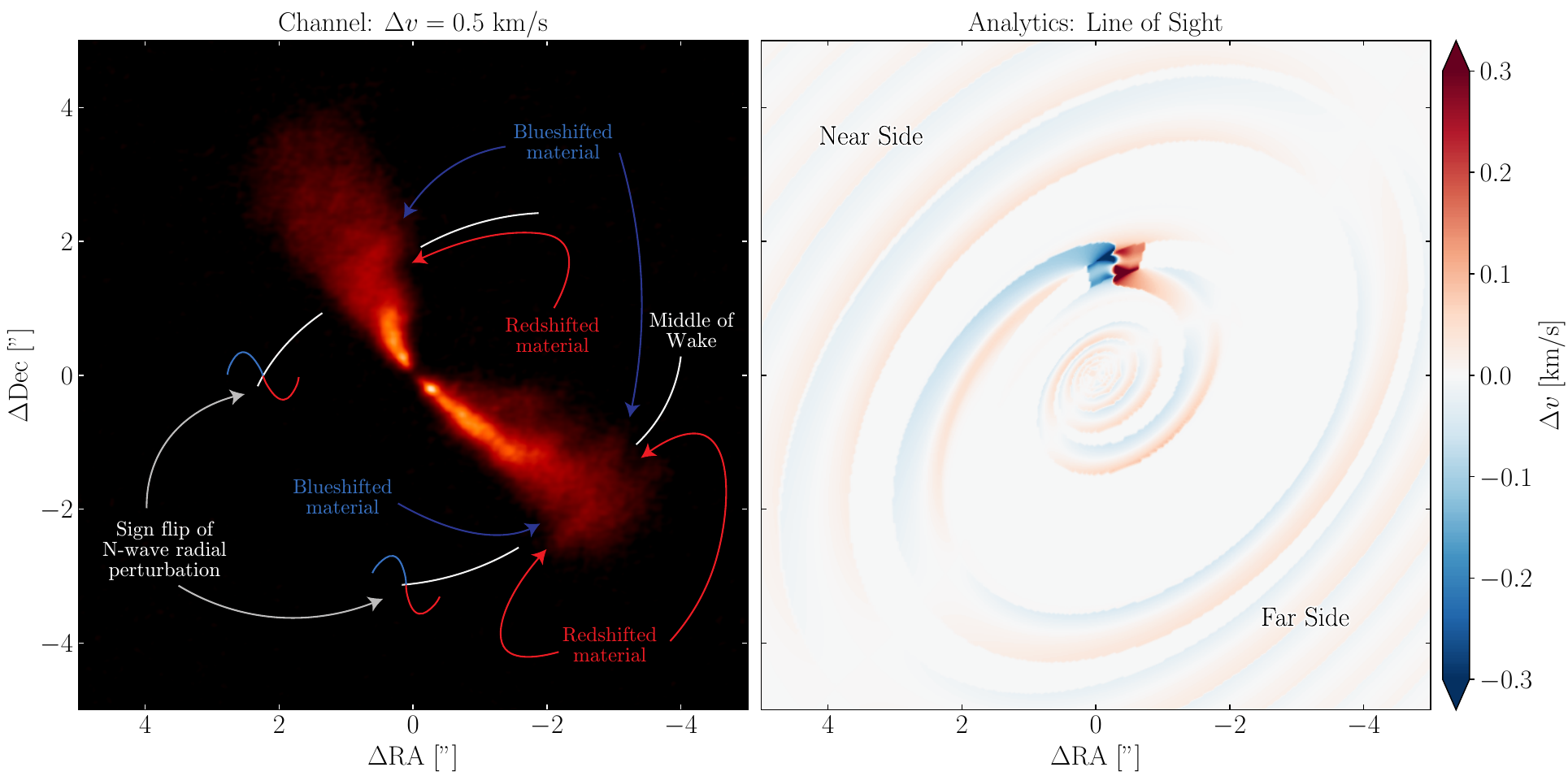}
\caption{Channel map in $^{12}$CO (2-1) line emission at $\Delta v=0.5 \, \mathrm{km / s}$ from systemic (left panel). We define positive velocity as moving away from the observer (redshifted) while negative velocity is towards the observer (blueshifted). Labels highlight redshifted and blueshifted material caused by the crossing of the planet wake. The sign flip over the semi-major axis indicates that the dominant velocity perturbation is radial, as expected for planet induced spiral density waves (right panel). }
\label{fig:lchan}
\end{figure*}

\begin{figure*}
\centering
\includegraphics[width=\textwidth]{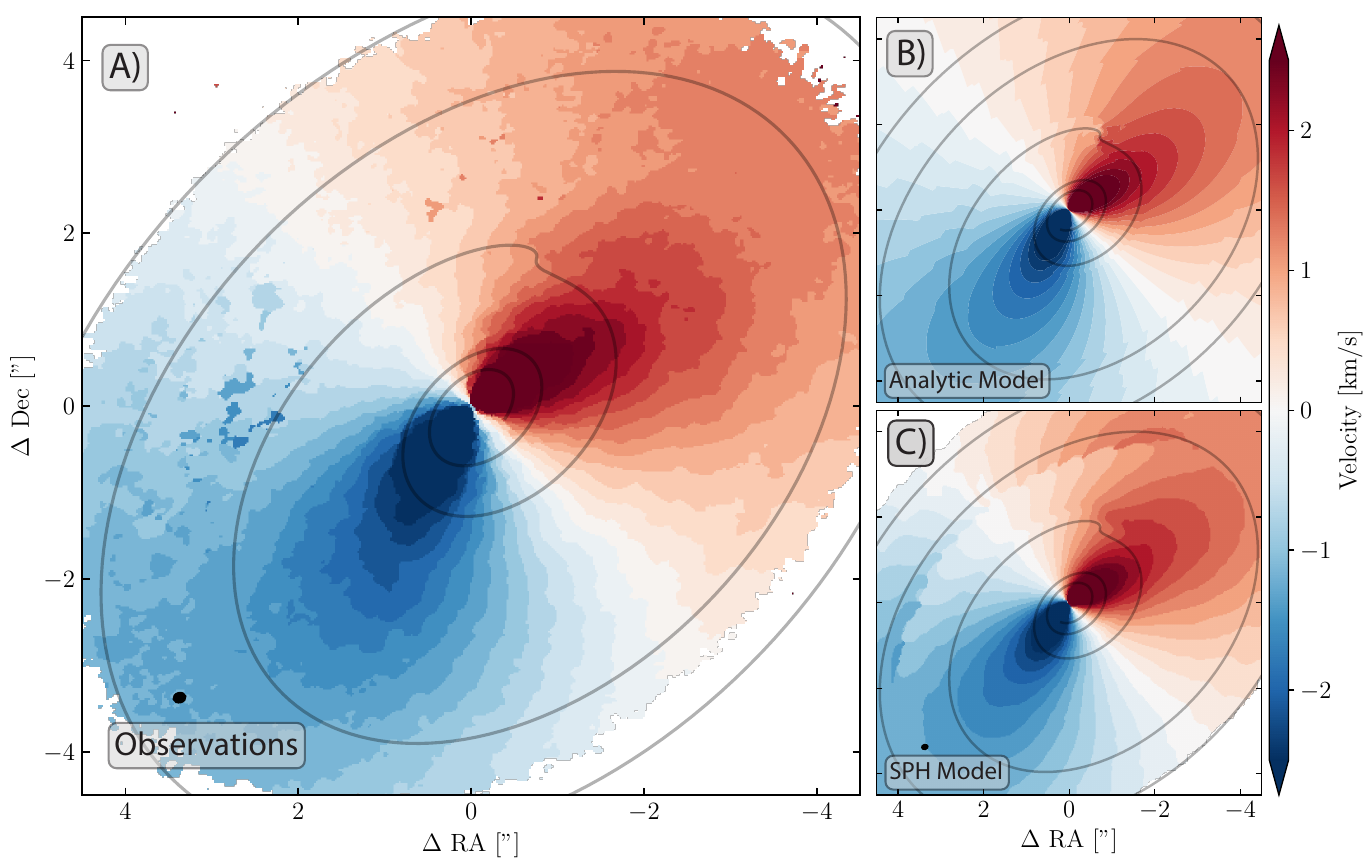}
\caption{Peak velocity map of $^{12}$CO (2-1) emission of HD~163296 from MAPS \protect{\citep{oberg2021}}  (panel A), the analytic model prediction (panel B), and the prediction from our SPH + radiative transfer model (panel C). The projected wake in the simulation and analytic models is shown by the solid line, and can be seen to trace a series of kinks mapping the wake over the whole disk. The same series of kinks is visible in the observations (panel A), revealing the wake from the planet in the data, almost perfectly corresponding to the projected wake from the models.
}
\label{fig:moment9}
\end{figure*}

\begin{figure*}
\centering
\includegraphics[width=0.8\textwidth]{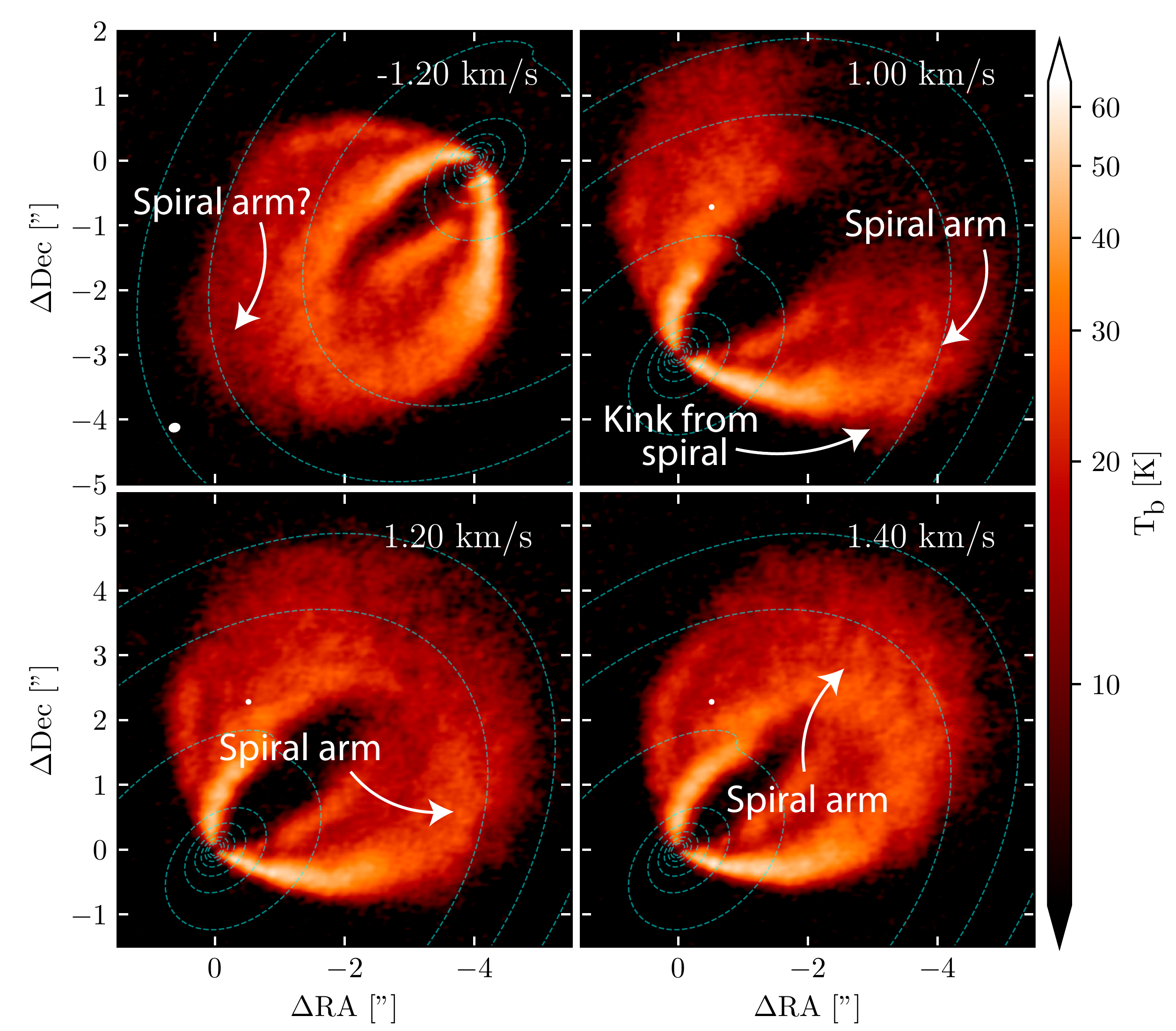}
\caption{A closer view of a selection of channels containing a faint spiral arm in CO emission. The location of the spiral is consistent with the projected location of the outer wake of the embedded protoplanet (dotted cyan line).} \label{fig:spiral}
\end{figure*}

\section{Results}

Figure \ref{fig:channels} displays a selection of CO channel maps of the disk around HD~163296 (subplot A) along with analogous channels from our SPH model (subplot B). We have annotated several features of interest. 
First, in the top left panel one may observe the velocity kink associated with an embedded planet, first reported by \cite{pinte2018}. In the third panel of the top row we highlight a secondary kink on the far side of the disk which is at a larger radial separation from the central star than the embedded protoplanet. 

We find the same structure in the SPH model (third panel from left in top half of subplot B). This is the outer wake of the planet. Interestingly this kink was predicted by the original SPH simulations by \cite{pinte2018} (see their Figure 5), and in retrospect can be seen in the original data used by \citet{pinte2018} to detect the embedded protoplanet. 

In subplot A of Figure \ref{fig:channels} we have labeled a faint kink seen on the lower surface of the disk (bottom row, second from left). An analogous feature is also seen in our SPH model. This is the outer wake of the planet seen on the lower surface of the disk. We plot the projected location of the outer planet wake on the lower surface to demonstrate this (magenta dashed line). Our SPH model suggests we should also see the spiral kink on the upper surface of the disk. Although there is a hint of this in the observations, the contrast is lower in this location with respect to the mid-plane and the lower surface.

Opaque crosses in Figure \ref{fig:channels} denote several additional kinks visible in the channel maps. Not all of these kinks are associated with the planet wake generated by the embedded protoplanet (see Section~\ref{sec:buoyancy}). The height of the CO emitting layer is lower in our model than in the observations. Increasing the disk aspect ratio would correct this, however the resulting planet wake would not be tightly wound enough as in the observations. The most likely reason for this discrepancy is a slight difference between the temperature profile assumed in the SPH model (which sets the wake propagation) and the temperature profile determined self-consistently in the radiative transfer by {\sc mcfost} (which sets the CO emitting layer). A proper treatment of the disk thermal structure in the SPH simulation is required to better understand the propagation of the wake and the disk scale height. We discuss the implications of this more in Section \ref{sec:limitations}.

Figure \ref{fig:analytics} shows the velocity perturbations predicted for a 3 M$_\textrm{J}$ planet in the \citetalias{bollati2021} analytic model. Left and middle panels show the velocity contribution in the radial and azimuthal directions, respectively. Comparing the left and middle panel we see that near the planet the azimuthal components dominate, as also seen in the supplementary material of \cite{pinte2019}, while far from the planet the radial motions dominate. That $\Delta v_\phi \ll \Delta v_r$ in the wake was mathematically proved by \citet[][Equations A4 and A8, respectively]{rafikov2002} so this is no surprise. The azimuthal contribution is only important within $\sim 10$ au of the planet. Figure \ref{fig:analytics} (right panel) shows the projection of the velocity perturbations on to the surface of the CO emitting layer in HD~163296. Since the perturbations are mostly radial, their projected amplitude decreases substantially along the semi-major axis of the disk where these perturbations are perpendicular to the line-of-sight.

Figure~\ref{fig:lchan} highlights the perturbations caused by the planet wake crossing the $\Delta v = 0.2$ km/s channel in the observations. Labels in the figure highlight the sign flip over the semi-major axis, as expected for radial dominated motion (see above). More subtle but visible nonetheless is the $N$-wave structure transverse to the wake predicted from the analytics (\citealt{rafikov2002}, \citetalias{bollati2021}).

Figure~\ref{fig:moment9} compares the peak velocity map of HD~163296 to the semi-analytic model (top right) and to our SPH + radiative transfer model (bottom right). 
Perturbations from Keplerian rotation, corresponding to `kinks' in the channel maps, are evident in the observations (panel A). In particular, the channelization of the data (in 200 m\,s$^{-1}$ bins) highlights that the isovelocity levels are distorted from the typical Keplerian butterfly pattern along the planet wake. These perturbations coincide with those predicted from the semi-analytic theory in the right panel of Figure \ref{fig:analytics} and in the predicted velocity map in Panel B. Similar perturbations are also seen in the SPH model (panel C) in agreement with the analytic model. Plotting the corresponding spiral wake from Eq.~\ref{eq:wake} in all three plots (solid line) \emph{reveals the extended planetary wake in the observational data}.

A final bonus is that the spiral arm is not just seen as velocity perturbations in the observations, but also in intensity.  Figure \ref{fig:spiral} demonstrates this, showing selected channels from the data. Beginning with the top right panel, one may first observe the velocity kink associated with the spiral. The label indicates where the spiral arm can also be seen in intensity. The spiral continues in to the preceding channels in the bottom left and right panels. The spiral is co-located with the predicted location of the outer planet wake generated by the embedded protoplanet, as shown by the dotted cyan line.

\section{Discussion}\label{sec:disc}
We have demonstrated that several of the recently reported spiral structures in HD~163296 \citep{teague2021} are due to planet wakes generated by the embedded protoplanet \citep{pinte2018}. These results provide the first direct confirmation of planetary spirals generated by Lindblad resonances \citep{goldreich1979, goldreich1980, ogilvie2002, goodman2001,rafikov2002}. We also show that the velocity damping prescription adopted by \citetalias{bollati2021} is not required. That is, the planet wake indeed induces secondary kinks in velocity channels that extend far from the planet location.

\subsection{Density waves or buoyancy spirals?}\label{sec:buoyancy}

 We have demonstrated that kinematic perturbations from planetary companions are more extended than previously assumed (\citetalias{bollati2021}, \citealt{teague2021}). The extended velocity perturbation associated with the embedded planet was also reported by \cite{teague2021} after subtraction of a flared Keplerian disk model. In their work, they suggest this is evidence of buoyancy spirals \citep{zhu2012b, bae2021} being excited by the embedded planet. Figure~\ref{fig:analytics} and panel B of Figure~\ref{fig:moment9} demonstrate that buoyancy spirals are not required to explain the secondary kinks \citepalias{bollati2021}.
 
However, several additional substructures are seen in the channels of Figure \ref{fig:channels} that do not appear to be associated with the planet wake of the embedded protoplanet. These features were also reported by \cite{teague2021}. We labeled these features with semi-opaque crosses in Figure \ref{fig:channels}. These additional features could possibly be attributed due to buoyancy resonances from the embedded protoplanet. Buoyancy spirals are expected to produce the greatest perturbations close to the planet \citep{bae2021}, and with a smaller opening angle. An alternative possibility is that these additional features arise from a second companion at a smaller radial separation, as suggested by \citet{teague2018}. \cite{pinte2020} reported the presence of a localized kink within the outer most gap of the continuum ring and interpreted this as a possible signature from an embedded planet. \cite{teague2021} also found evidence of a velocity perturbation in this region. 
If such a planet does exist, we may be witnessing the outer wake of this planet.

Figure~\ref{fig:lchan} confirms our hypothesis that the spirals are density waves rather than buoyancy spirals. Density waves are dominated by radial motions \citep{rafikov2002}, whereas buoyancy waves produce mostly vertical motions \citep{bae2021}. As seen in Figure~\ref{fig:analytics} (right panel) radial velocity perturbations change sign along the semi-major axis, while vertical motions would not. Figure~\ref{fig:lchan} shows that the sign flip expected for radial dominated motions is present in the observational data. Another approach would be to look for this sign change in residual maps \citep{teague2021, discminer}, but we found that the sign flip measured this way is easily confused even in residuals from simulations, as it depends sensitively on the background disk model subtracted (see Appendix \ref{sec:appendix}). 

\subsection{Estimating the planet mass}\label{sec:mass}
The observed planet wake in HD~163296 suggests a more robust way to estimate properties of the embedded planet and the protoplanetary disk. As suggested by \citetalias{bollati2021}, the amplitude of the velocity kink generated by the spiral as it crosses a channel is related to the perturber mass. The problem is that the velocity perturbation, and thus the kink amplitude, is related to the embedded planet mass in units of thermal mass as seen in equation~9 of \citetalias{bollati2021}. The thermal mass is given by \citep{goodman2001}
\begin{equation}
    M_\mathrm{th} = \frac{2}{3} \left( \frac{h_p}{r_p} \right)^3 M_\star,
\end{equation}
and depends sensitively on the disk thermal structure. Thus, to measure the absolute planet mass one must break the degeneracy between the perturbation amplitude and the disk thermal structure (i.e. $h_p/r_p$) \citep[see][]{bae2018b}. 

While good constraints on the disk temperature structure are possible from the CO emission \citep[e.g.][]{pinte2018b}, this does not trace the mid-plane gas temperature or the pressure scale height at the planet location which is needed to measure the thermal mass. The CO emission between the upper and lower surface may allow for a direct measurement of the mid-plane gas temperature \citep{Dullemond2020}, but the assumption of optically thick emission (for which T$_\mathrm{b}$ = T$_\mathrm{gas}$) is not guaranteed to be valid due to freeze-out of the molecules on the dust grains. Mapping the spiral wake provides a novel way to break this degeneracy and determine the perturber mass using only the disk kinematics. As seen in equation \eqref{eq:wake}, the propagation of the spiral wake depends on the disk aspect ratio. Thus by measuring both the amplitude of the kinks close to the planet (to determine the mass in units of the thermal mass) and fitting the wake shape to the location of the spiral-induced velocity kinks (to measure $h/r$ as a function of $r$), one can in principle directly determine the mass of the embedded planet using the non-linear wake propagation theory. \cite{Rabago2021} have also suggested constraining the planet mass using the maximum kink velocity.

We have shown that the shape of the planet wake in the data can be used to constrain the disk aspect ratio. This is needed for an accurate estimate of the thermal mass. Fitting the planet mass itself would require simultaneous fitting of the kink amplitude, the wake shape, and the emitting surface, which is a complex procedure and beyond the scope of this Letter.

\subsection{Limitations of the models}\label{sec:limitations}

\citet{zhu2015} and \citet{bae2018a} found that the wake predicted by Equation~\eqref{eq:wake} underestimates the pitch angle of the spiral arm for large planet masses. \cite{cimerman2021} also found that the non-linear planet wake propagation theory overpredicts the decay of the density wave. The perturbation amplitude near the planet may also be wrong for planet masses exceeding the thermal mass \citep{rafikov2002}. This is the reason we compared to both simulations and analytics, as the simulations do not have this limitation. Since we find similar results in both cases, this suggests the effects on the observed wake are limited.

In both the simulations and analytic models we assumed a vertically isothermal temperature profile, which is not strictly true in the disk around HD~163296 \citep{rosenfeld2013,deGregorioMonsalvo2013}. Temperature stratification would mean that the sound wave launched by the planet would propagate faster in the (hotter) upper layers of the disc, and result in spirals with a larger pitch angle \citep{juhasz2018}. Additional spirals due to buoyancy can also arise when there is a vertical temperature gradient \citep{bae2021}. However, as seen in Figure 10 of \citet{law2021} the temperature stratification is not significant in the outer regions of the disk up to several scale heights, which is also the region we are principally concerned with in this Letter. Additionally, the supplementary material of \citet{pinte2019} shows similar morphology for the wake generated in both isothermal and stratified disks, although they only compared the regions close to the planet.

Additionally, wake propagation in 3D deviates from the linear wave theory of equation~\eqref{eq:wake}. \cite{zhu2015} showed that the pitch angle of the outer planetary wake is smaller (i.e. the spiral is more tightly wound) with increasing height above the disk mid-plane. This is taken into account in our SPH models, but is not when we project the wake onto the CO emitting layer. This effect will somewhat counteract the faster sound speed of the CO emitting layer relative to the mid-plane. However given the thermal stratification is not significant in the outer disk this may be negligible. Additionally shock heating also changes the wake shape and propagation \citep{zhu2015}. A more robust treatment of the disk thermodynamics (e.g. accounting for radiative and shock heating) is required to explore this issue but is beyond the scope of this Letter.

\section{Summary}
We demonstrate that recently reported spiral structure in HD~163296 \citep{teague2021} is the outer planetary wake generated by an embedded protoplanet \citep{pinte2018}, providing the first confirmation of planetary wakes generated by Lindblad resonances.

\section*{Acknowledgments}
We thank Richard Teague and Andr\'es Izquierdo for useful discussion. This research used resources provided by the Los Alamos National Laboratory Institutional Computing Program, which is supported by the U.S. Department of Energy National Nuclear Security Administration under Contract No. 89233218CNA000001. JC acknowledges the support from LANL/LDRD program (approved for public release as LA-UR-21-31084). DJP and CP acknowledge funding from the Australian Research Council via FT130100034, DP180104235 and FT170100040. BJN is supported by an Australian Government Research Training Program (RTP) Scholarship. This project has received funding from the European Union’s Horizon 2020 research and innovation program under the Marie Sklodowska-Curie Grant agreement 823823 (DUSTBUSTERS). We used {\sc plonk} \citep{plonk2019} which utilizes functions written in {\sc splash} \citep{price2007}.

\bibliography{paper}{}

\begin{thebibliography}{}
\expandafter\ifx\csname natexlab\endcsname\relax\def\natexlab#1{#1}\fi
\providecommand{\url}[1]{\href{#1}{#1}}
\providecommand{\dodoi}[1]{doi:~\href{http://doi.org/#1}{\nolinkurl{#1}}}
\providecommand{\doeprint}[1]{\href{http://ascl.net/#1}{\nolinkurl{http://ascl.net/#1}}}
\providecommand{\doarXiv}[1]{\href{https://arxiv.org/abs/#1}{\nolinkurl{https://arxiv.org/abs/#1}}}

\bibitem[{{Andrews} {et~al.}(2018){Andrews}, {Huang}, {P{\'e}rez}, {Isella},
  {Dullemond}, {Kurtovic}, {Guzm{\'a}n}, {Carpenter}, {Wilner}, {Zhang}, {Zhu},
  {Birnstiel}, {Bai}, {Benisty}, {Hughes}, {{\"O}berg}, \&
  {Ricci}}]{andrews2018}
{Andrews}, S.~M., {Huang}, J., {P{\'e}rez}, L.~M., {et~al.} 2018, \apjl, 869,
  L41, \dodoi{10.3847/2041-8213/aaf741}

\bibitem[{{Bae} {et~al.}(2021){Bae}, {Teague}, \& {Zhu}}]{bae2021}
{Bae}, J., {Teague}, R., \& {Zhu}, Z. 2021, \apj, 912, 56,
  \dodoi{10.3847/1538-4357/abe45e}

\bibitem[{{Bae} \& {Zhu}(2018{\natexlab{a}})}]{bae2018b}
{Bae}, J., \& {Zhu}, Z. 2018{\natexlab{a}}, \apj, 859, 119,
  \dodoi{10.3847/1538-4357/aabf93}

\bibitem[{{Bae} \& {Zhu}(2018{\natexlab{b}})}]{bae2018a}
---. 2018{\natexlab{b}}, \apj, 859, 118, \dodoi{10.3847/1538-4357/aabf8c}

\bibitem[{{Bate} {et~al.}(1995){Bate}, {Bonnell}, \& {Price}}]{bate1995}
{Bate}, M.~R., {Bonnell}, I.~A., \& {Price}, N.~M. 1995, \mnras, 277, 362,
  \dodoi{10.1093/mnras/277.2.362}

\bibitem[{{Bollati} {et~al.}(2021){Bollati}, {Lodato}, {Price}, \&
  {Pinte}}]{bollati2021}
{Bollati}, F., {Lodato}, G., {Price}, D.~J., \& {Pinte}, C. 2021, \mnras, 504,
  5444, \dodoi{10.1093/mnras/stab1145}

\bibitem[{{Cimerman} \& {Rafikov}(2021)}]{cimerman2021}
{Cimerman}, N.~P., \& {Rafikov}, R.~R. 2021, \mnras, 508, 2329,
  \dodoi{10.1093/mnras/stab2652}

\bibitem[{{Czekala} {et~al.}(2021){Czekala}, {Loomis}, {Teague}, {Booth},
  {Huang}, {Cataldi}, {Ilee}, {Law}, {Walsh}, {Bosman}, {Guzm{\'a}n}, {Le Gal},
  {{\"O}berg}, {Yamato}, {Aikawa}, {Andrews}, {Bae}, {Bergin}, {Bergner},
  {Cleeves}, {Kurtovic}, {M{\'e}nard}, {Nomura}, {P{\'e}rez}, {Qi}, {Schwarz},
  {Tsukagoshi}, {Waggoner}, {Wilner}, \& {Zhang}}]{MAPSII2021}
{Czekala}, I., {Loomis}, R.~A., {Teague}, R., {et~al.} 2021, arXiv e-prints,
  arXiv:2109.06188.
\newblock \doarXiv{2109.06188}

\bibitem[{{de Gregorio-Monsalvo} {et~al.}(2013){de Gregorio-Monsalvo},
  {M{\'e}nard}, {Dent}, {Pinte}, {L{\'o}pez}, {Klaassen}, {Hales},
  {Cort{\'e}s}, {Rawlings}, {Tachihara}, {Testi}, {Takahashi}, {Chapillon},
  {Mathews}, {Juhasz}, {Akiyama}, {Higuchi}, {Saito}, {Nyman}, {Phillips},
  {Rod{\'o}n}, {Corder}, \& {Van Kempen}}]{deGregorioMonsalvo2013}
{de Gregorio-Monsalvo}, I., {M{\'e}nard}, F., {Dent}, W., {et~al.} 2013, \aap,
  557, A133, \dodoi{10.1051/0004-6361/201321603}

\bibitem[{{Dullemond} {et~al.}(2020){Dullemond}, {Isella}, {Andrews},
  {Skobleva}, \& {Dzyurkevich}}]{Dullemond2020}
{Dullemond}, C.~P., {Isella}, A., {Andrews}, S.~M., {Skobleva}, I., \&
  {Dzyurkevich}, N. 2020, \aap, 633, A137, \dodoi{10.1051/0004-6361/201936438}

\bibitem[{{Goldreich} \& {Tremaine}(1979)}]{goldreich1979}
{Goldreich}, P., \& {Tremaine}, S. 1979, \apj, 233, 857, \dodoi{10.1086/157448}

\bibitem[{{Goldreich} \& {Tremaine}(1980)}]{goldreich1980}
---. 1980, \apj, 241, 425, \dodoi{10.1086/158356}

\bibitem[{{Goodman} \& {Rafikov}(2001)}]{goodman2001}
{Goodman}, J., \& {Rafikov}, R.~R. 2001, \apj, 552, 793, \dodoi{10.1086/320572}

\bibitem[{{Grady} {et~al.}(2000){Grady}, {Devine}, {Woodgate}, {Kimble},
  {Bruhweiler}, {Boggess}, {Linsky}, {Plait}, {Clampin}, \&
  {Kalas}}]{grady2000}
{Grady}, C.~A., {Devine}, D., {Woodgate}, B., {et~al.} 2000, \apj, 544, 895,
  \dodoi{10.1086/317222}

\bibitem[{{Huang} {et~al.}(2018){Huang}, {Andrews}, {Dullemond}, {Isella},
  {P{\'e}rez}, {Guzm{\'a}n}, {{\"O}berg}, {Zhu}, {Zhang}, {Bai}, {Benisty},
  {Birnstiel}, {Carpenter}, {Hughes}, {Ricci}, {Weaver}, \&
  {Wilner}}]{jhuang2018}
{Huang}, J., {Andrews}, S.~M., {Dullemond}, C.~P., {et~al.} 2018, \apjl, 869,
  L42, \dodoi{10.3847/2041-8213/aaf740}

\bibitem[{{Izquierdo} {et~al.}(2021){Izquierdo}, {Facchini}, {Rosotti}, {van
  Dishoeck}, \& {Testi}}]{discminer}
{Izquierdo}, A.~F., {Facchini}, S., {Rosotti}, G.~P., {van Dishoeck}, E.~F., \&
  {Testi}, L. 2021, arXiv e-prints, arXiv:2111.06367.
\newblock \doarXiv{2111.06367}

\bibitem[{{Jorsater} \& {van Moorsel}(1995)}]{jvm1995}
{Jorsater}, S., \& {van Moorsel}, G.~A. 1995, \aj, 110, 2037,
  \dodoi{10.1086/117668}

\bibitem[{{Juh{\'a}sz} \& {Rosotti}(2018)}]{juhasz2018}
{Juh{\'a}sz}, A., \& {Rosotti}, G.~P. 2018, \mnras, 474, L32,
  \dodoi{10.1093/mnrasl/slx182}

\bibitem[{{Law} {et~al.}(2021){Law}, {Teague}, {Loomis}, {Bae}, {{\"O}berg},
  {Czekala}, {Andrews}, {Aikawa}, {Alarc{\'o}n}, {Bergin}, {Bergner}, {Booth},
  {Bosman}, {Calahan}, {Cataldi}, {Cleeves}, {Furuya}, {Guzm{\'a}n}, {Huang},
  {Ilee}, {Le Gal}, {Liu}, {Long}, {M{\'e}nard}, {Nomura}, {P{\'e}rez}, {Qi},
  {Schwarz}, {Soto}, {Tsukagoshi}, {Yamato}, {van't Hoff}, {Walsh}, {Wilner},
  \& {Zhang}}]{law2021}
{Law}, C.~J., {Teague}, R., {Loomis}, R.~A., {et~al.} 2021, arXiv e-prints,
  arXiv:2109.06217.
\newblock \doarXiv{2109.06217}

\bibitem[{{Mentiplay}(2019)}]{plonk2019}
{Mentiplay}, D. 2019, The Journal of Open Source Software, 4, 1884,
  \dodoi{10.21105/joss.01884}

\bibitem[{{Oberg} {et~al.}(2021{\natexlab{a}}){Oberg}, {Guzman}, {Walsh},
  {Aikawa}, {Bergin}, {Law}, {Loomis}, {Alarcon}, {Andrews}, {Bae}, {Bergner},
  {Boehler}, {Booth}, {Bosman}, {Calahan}, {Cataldi}, {Cleeves}, {Czekala},
  {Furuya}, {Huang}, {Ilee}, {Kurtovic}, {Le Gal}, {Liu}, {Long}, {Menard},
  {Nomura}, {Perez}, {Qi}, {Schwarz}, {Sierra}, {Teague}, {Tsukagoshi},
  {Yamato}, {van 't Hoff}, {Waggoner}, {Wilner}, \& {Zhang}}]{oberg2021}
{Oberg}, K.~I., {Guzman}, V.~V., {Walsh}, C., {et~al.} 2021{\natexlab{a}},
  arXiv e-prints, arXiv:2109.06268.
\newblock \doarXiv{2109.06268}

\bibitem[{{Oberg} {et~al.}(2021{\natexlab{b}}){Oberg}, {Guzman}, {Walsh},
  {Aikawa}, {Bergin}, {Law}, {Loomis}, {Alarcon}, {Andrews}, {Bae}, {Bergner},
  {Boehler}, {Booth}, {Bosman}, {Calahan}, {Cataldi}, {Cleeves}, {Czekala},
  {Furuya}, {Huang}, {Ilee}, {Kurtovic}, {Le Gal}, {Liu}, {Long}, {Menard},
  {Nomura}, {Perez}, {Qi}, {Schwarz}, {Sierra}, {Teague}, {Tsukagoshi},
  {Yamato}, {van 't Hoff}, {Waggoner}, {Wilner}, \& {Zhang}}]{MAPSI2021}
---. 2021{\natexlab{b}}, arXiv e-prints, arXiv:2109.06268.
\newblock \doarXiv{2109.06268}

\bibitem[{{Ogilvie} \& {Lubow}(2002)}]{ogilvie2002}
{Ogilvie}, G.~I., \& {Lubow}, S.~H. 2002, \mnras, 330, 950,
  \dodoi{10.1046/j.1365-8711.2002.05148.x}

\bibitem[{{Pinte} {et~al.}(2009){Pinte}, {Harries}, {Min}, {Watson},
  {Dullemond}, {Woitke}, {M{\'e}nard}, \& {Dur{\'a}n-Rojas}}]{pinte2009}
{Pinte}, C., {Harries}, T.~J., {Min}, M., {et~al.} 2009, \aap, 498, 967,
  \dodoi{10.1051/0004-6361/200811555}

\bibitem[{{Pinte} {et~al.}(2006){Pinte}, {M{\'e}nard}, {Duch{\^e}ne}, \&
  {Bastien}}]{pinte2006}
{Pinte}, C., {M{\'e}nard}, F., {Duch{\^e}ne}, G., \& {Bastien}, P. 2006, \aap,
  459, 797, \dodoi{10.1051/0004-6361:20053275}

\bibitem[{{Pinte} {et~al.}(2018{\natexlab{a}}){Pinte}, {Price}, {M{\'e}nard},
  {Duch{\^e}ne}, {Dent}, {Hill}, {de Gregorio-Monsalvo}, {Hales}, \&
  {Mentiplay}}]{pinte2018}
{Pinte}, C., {Price}, D.~J., {M{\'e}nard}, F., {et~al.} 2018{\natexlab{a}},
  \apjl, 860, L13, \dodoi{10.3847/2041-8213/aac6dc}

\bibitem[{{Pinte} {et~al.}(2018{\natexlab{b}}){Pinte}, {M{\'e}nard},
  {Duch{\^e}ne}, {Hill}, {Dent}, {Woitke}, {Maret}, {van der Plas}, {Hales},
  {Kamp}, {Thi}, {de Gregorio-Monsalvo}, {Rab}, {Quanz}, {Avenhaus}, {Carmona},
  \& {Casassus}}]{pinte2018b}
{Pinte}, C., {M{\'e}nard}, F., {Duch{\^e}ne}, G., {et~al.} 2018{\natexlab{b}},
  \aap, 609, A47, \dodoi{10.1051/0004-6361/201731377}

\bibitem[{{Pinte} {et~al.}(2019){Pinte}, {van der Plas}, {M{\'e}nard}, {Price},
  {Christiaens}, {Hill}, {Mentiplay}, {Ginski}, {Choquet}, {Boehler},
  {Duch{\^e}ne}, {Perez}, \& {Casassus}}]{pinte2019}
{Pinte}, C., {van der Plas}, G., {M{\'e}nard}, F., {et~al.} 2019, Nature
  Astronomy, 3, 1109, \dodoi{10.1038/s41550-019-0852-6}

\bibitem[{{Pinte} {et~al.}(2020){Pinte}, {Price}, {M{\'e}nard}, {Duch{\^e}ne},
  {Christiaens}, {Andrews}, {Huang}, {Hill}, {van der Plas}, {Perez}, {Isella},
  {Boehler}, {Dent}, {Mentiplay}, \& {Loomis}}]{pinte2020}
{Pinte}, C., {Price}, D.~J., {M{\'e}nard}, F., {et~al.} 2020, \apjl, 890, L9,
  \dodoi{10.3847/2041-8213/ab6dda}

\bibitem[{{Price}(2007)}]{price2007}
{Price}, D.~J. 2007, \pasa, 24, 159, \dodoi{10.1071/AS07022}

\bibitem[{{Price} {et~al.}(2018){Price}, {Wurster}, {Tricco}, {Nixon},
  {Toupin}, {Pettitt}, {Chan}, {Mentiplay}, {Laibe}, {Glover}, {Dobbs},
  {Nealon}, {Liptai}, {Worpel}, {Bonnerot}, {Dipierro}, {Ballabio}, {Ragusa},
  {Federrath}, {Iaconi}, {Reichardt}, {Forgan}, {Hutchison}, {Constantino},
  {Ayliffe}, {Hirsh}, \& {Lodato}}]{phantom2018}
{Price}, D.~J., {Wurster}, J., {Tricco}, T.~S., {et~al.} 2018, \pasa, 35, e031,
  \dodoi{10.1017/pasa.2018.25}

\bibitem[{{Rabago} \& {Zhu}(2021)}]{Rabago2021}
{Rabago}, I., \& {Zhu}, Z. 2021, \mnras, 502, 5325,
  \dodoi{10.1093/mnras/stab447}

\bibitem[{{Rafikov}(2002)}]{rafikov2002}
{Rafikov}, R.~R. 2002, \apj, 569, 997, \dodoi{10.1086/339399}

\bibitem[{{Rosenfeld} {et~al.}(2013){Rosenfeld}, {Andrews}, {Hughes}, {Wilner},
  \& {Qi}}]{rosenfeld2013}
{Rosenfeld}, K.~A., {Andrews}, S.~M., {Hughes}, A.~M., {Wilner}, D.~J., \&
  {Qi}, C. 2013, \apj, 774, 16, \dodoi{10.1088/0004-637X/774/1/16}

\bibitem[{{Setterholm} {et~al.}(2018){Setterholm}, {Monnier}, {Davies},
  {Kreplin}, {Kraus}, {Baron}, {Aarnio}, {Berger}, {Calvet}, {Cur{\'e}},
  {Kanaan}, {Kloppenborg}, {Le Bouquin}, {Millan-Gabet}, {Rubinstein}, {Sitko},
  {Sturmann}, {ten Brummelaar}, \& {Touhami}}]{Setterholm2018}
{Setterholm}, B.~R., {Monnier}, J.~D., {Davies}, C.~L., {et~al.} 2018, \apj,
  869, 164, \dodoi{10.3847/1538-4357/aaef2c}

\bibitem[{{Takeuchi} \& {Lin}(2002)}]{takeuchi2002}
{Takeuchi}, T., \& {Lin}, D.~N.~C. 2002, \apj, 581, 1344,
  \dodoi{10.1086/344437}

\bibitem[{Teague(2019)}]{eddy}
Teague, R. 2019, The Journal of Open Source Software, 4, 1220,
  \dodoi{10.21105/joss.01220}

\bibitem[{{Teague} {et~al.}(2018){Teague}, {Bae}, {Bergin}, {Birnstiel}, \&
  {Foreman-Mackey}}]{teague2018}
{Teague}, R., {Bae}, J., {Bergin}, E.~A., {Birnstiel}, T., \& {Foreman-Mackey},
  D. 2018, \apjl, 860, L12, \dodoi{10.3847/2041-8213/aac6d7}

\bibitem[{Teague \& Foreman-Mackey(2018)}]{bettermoments2018}
Teague, R., \& Foreman-Mackey, D. 2018, Zenodo, \dodoi{10.5281/zenodo.1419754}

\bibitem[{{Teague} {et~al.}(2021){Teague}, {Bae}, {Aikawa}, {Andrews},
  {Bergin}, {Bergner}, {Boehler}, {Booth}, {Bosman}, {Cataldi}, {Czekala},
  {Guzm{\'a}n}, {Huang}, {Ilee}, {Law}, {Le Gal}, {Long}, {Loomis},
  {M{\'e}nard}, {{\"O}berg}, {P{\'e}rez}, {Schwarz}, {Sierra}, {Walsh},
  {Wilner}, {Yamato}, \& {Zhang}}]{teague2021}
{Teague}, R., {Bae}, J., {Aikawa}, Y., {et~al.} 2021, arXiv e-prints,
  arXiv:2109.06218.
\newblock \doarXiv{2109.06218}

\bibitem[{{Weingartner} \& {Draine}(2001)}]{weingartner2001}
{Weingartner}, J.~C., \& {Draine}, B.~T. 2001, \apj, 548, 296,
  \dodoi{10.1086/318651}

\bibitem[{{Zhu} {et~al.}(2015){Zhu}, {Dong}, {Stone}, \& {Rafikov}}]{zhu2015}
{Zhu}, Z., {Dong}, R., {Stone}, J.~M., \& {Rafikov}, R.~R. 2015, \apj, 813, 88,
  \dodoi{10.1088/0004-637X/813/2/88}

\bibitem[{{Zhu} {et~al.}(2012){Zhu}, {Stone}, \& {Rafikov}}]{zhu2012b}
{Zhu}, Z., {Stone}, J.~M., \& {Rafikov}, R.~R. 2012, \apjl, 758, L42,
  \dodoi{10.1088/2041-8205/758/2/L42}

\end{thebibliography}
\bibliographystyle{aasjournal}


\appendix

\section{Residuals from Keplerian Rotation}\label{sec:appendix}

\begin{figure*}
\centering
\includegraphics[width=\textwidth]{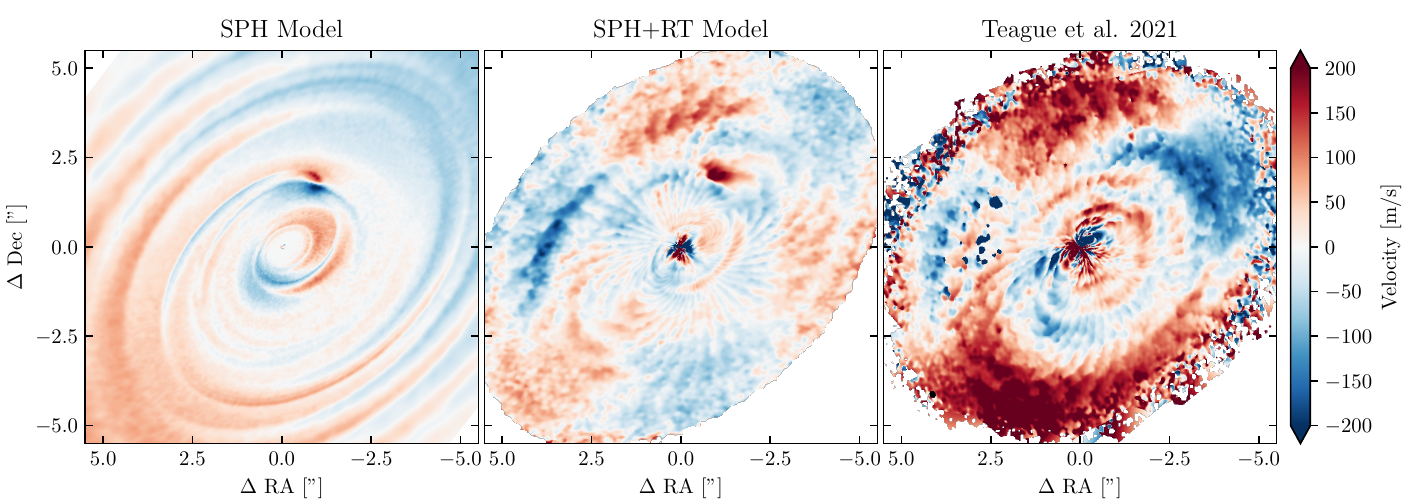}
\caption{Residuals from Keplerian rotation predicted from our SPH model (left panel), from our SPH model after radiative transfer (middle panel), and from the observations presented in \citet{teague2021}. Our SPH model shows radial, vertical, and azimuthal velocity deviations from circular Keplerian rotation. We used {\sc Eddy} to subtract a flared Keplerian disk model from out SPH+RT model. {\sc Eddy} does not recover the true residuals in our simulation because of errors in the stellar mass and CO emitting layer determination, but the residual map contains all the major features of the residual map produced in the same manner from the observations \citep[shown in Figure 1b of][]{teague2021}.} \label{fig:residuals}
\end{figure*}

In Figure \ref{fig:residuals} we plot the Keplerian velocity residuals from our SPH before and after radiative transfer (left and middle), along with the velocity residuals from \citet{teague2021}. We used the code {\sc bettermoments} to generate the velocity map of our SPH+RT model and used the quadratic method \citep{bettermoments2018}. We follow the fitting procedure outlined in Section 3.1 of \citet{teague2021} to fit our radiative transfer model and obtain a best fit CO emitting layer, stellar mass, and disk position angle. The left panel of Figure \ref{fig:residuals} shows the deviations from Keplerian rotation of the SPH particles close to this layer (accounting for their height above the disk mid-plane).

An issue with this approach is that {\sc Eddy} \citep{eddy}, when fed the synthetic moment map from our SPH simulation (middle panel in Figure~\ref{fig:residuals}) fits a stellar mass of 1.84 M$_\odot$, whereas the mass of the central object in our model was 1.9M$_\odot$. This difference is likely due to the gas pressure support not being accounted for in the fitting model. However the resulting error in the residuals from the incorrect stellar mass are $v_{\rm res} = \sqrt{{G\Delta M_*}/{r_{\rm p}}} \approx 460\,{\rm m}/{\rm s}$ with $\Delta M_* = 0.06$ M$_\odot$ and $r_{\rm p} = 250$ au --- on the same order as the residuals of $\approx 300 {\rm m}/{\rm s}$ from the planet wake (Figure~\ref{fig:analytics}). These rather minor errors in the stellar mass determination produce residuals which do not reflect those in the SPH model. A similar effect could also be generated by small changes in the CO emitting layer. Since these small differences in model parameters produce velocity deviations on the order of the velocity deviations induced by the kink, they can produce a residual map that hides the expected kink signature seen the left panel of Figure \ref{fig:residuals} and right panel of Figure \ref{fig:lchan}. This also explains why residual maps produced from either our simulations or the observations \citep{teague2021} do not obviously show the sign flip across the major axis expected for radial motions.

\end{document}